\documentclass[12pt]{article}
\usepackage{amssymb}

\def\hybrid{\topmargin 0pt      \oddsidemargin 0pt
        \headheight 0pt \headsep 0pt
        \voffset=-0.5cm
        \hoffset=-0.25in
        \textwidth 6.75in
        \textheight 9.5in       
        \marginparwidth 0.0in
        \parskip 5pt plus 1pt   \jot = 1.5ex}
\catcode`\@=11
\def\marginnote#1{}

\newcount\hour
\newcount\minute
\newtoks\amorpm
\hour=\time\divide\hour by60 \minute=\time{\multiply\hour by60
\global\advance\minute by-\hour}
\edef\standardtime{{\ifnum\hour<12 \global\amorpm={am}%
        \else\global\amorpm={pm}\advance\hour by-12 \fi
        \ifnum\hour=0 \hour=12 \fi
        \number\hour:\ifnum\minute<10 0\fi\number\minute\the\amorpm}}
\edef\militarytime{\number\hour:\ifnum\minute<10 0\fi\number\minute}

\def\draftlabel#1{{\@bsphack\if@filesw {\let\thepage\relax
   \xdef\@gtempa{\write\@auxout{\string
      \newlabel{#1}{{\@currentlabel}{\thepage}}}}}\@gtempa
   \if@nobreak \ifvmode\nobreak\fi\fi\fi\@esphack}
        \gdef\@eqnlabel{#1}}
\def\@eqnlabel{}
\def\@vacuum{}
\def\draftmarginnote#1{\marginpar{\raggedright\scriptsize\tt#1}}
\def\draftlabel#1{{\@bsphack\if@filesw {\let\thepage\relax
   \xdef\@gtempa{\write\@auxout{\string
      \newlabel{#1}{{\@currentlabel}{\thepage}}}}}\@gtempa
   \if@nobreak \ifvmode\nobreak\fi\fi\fi\@esphack}
        \gdef\@eqnlabel{#1}}
\def\@eqnlabel{}
\def\@vacuum{}
\def\draftmarginnote#1{\marginpar{\raggedright\scriptsize\tt#1}}

\def\draft{\oddsidemargin -.5truein
        \def\@oddfoot{\sl preliminary draft \hfil
        \rm\thepage\hfil\sl\today\quad\militarytime}
        \let\@evenfoot\@oddfoot \overfullrule 3pt
        \let\label=\draftlabel
        \let\marginnote=\draftmarginnote
   \def\@eqnnum{(\theequation)\rlap{\kern\marginparsep\tt\@eqnlabel}%
\global\let\@eqnlabel\@vacuum}  }

\def\numberbysection{\@addtoreset{equation}{section}
        \def\theequation{\thesection.\arabic{equation}}}

\def\underline#1{\relax\ifmmode\@@underline#1\else
        $\@@underline{\hbox{#1}}$\relax\fi}

\def\titlepage{\@restonecolfalse\if@twocolumn\@restonecoltrue\onecolumn
     \else \newpage \fi \thispagestyle{empty}\c@page\z@
        \def\thefootnote{\fnsymbol{footnote}} }

\def\endtitlepage{\if@restonecol\twocolumn \else  \fi
        \def\thefootnote{\arabic{footnote}}
        \setcounter{footnote}{0}}  

\numberbysection
 \hybrid

\newcounter{mo}

\newcommand{\tr}{{\rm tr}}
\newcommand{\ti}[1]{\tilde{#1}}

\newcommand{\vf}{\varphi}
\newcommand{\al}{\alpha}
\newcommand{\ka}{\kappa}
\newcommand{\be}{\beta}
\newcommand{\ga}{\gamma}
\newcommand{\om}{\omega}
\newcommand{\vth}{\vartheta}

\newcommand{\Mat}{ {\rm Mat}(N,\mathbb C) }
\newcommand{\Matm}{ {\rm Mat}(M,\mathbb C) }
\newcommand{\mR}{\mathfrak R}

\newcommand{\mC}{\mathbb C}
\newcommand{\mZ}{\mathbb Z}
\newcommand{\bS}{{\mathbb{S}}}

\def\beq{\begin{equation}}
\def\eq{\end{equation}}
\def\p{\partial}

\newcommand{\tal}{{\ti\alpha}}
\newcommand{\tbe}{{\ti\be}}
\newcommand{\tga}{{\ti\gamma}}
\newcommand{\tom}{{\ti\omega}}
\newcommand{\tT}{{\ti T}}

\def\res{\mathop{\hbox{Res}}\limits}

\begin{document}

\setcounter{page}{1}

\date{}
\date{}
\vspace{50mm}


\begin{center}
\vspace{3mm}
%
{\LARGE{Relativistic elliptic matrix tops}}
 \\ \vspace{4mm}
{\LARGE{and finite Fourier transformations}}
%
\\
\vspace{12mm} {\large  {A. Zotov} }\\
 \vspace{8mm}

  {\small{\sf Steklov Mathematical Institute of Russian
Academy of Sciences,\\ Gubkina str. 8, Moscow, 119991,  Russia}}
\end{center}

\begin{center}\footnotesize{{\rm E-mail:}{\rm\ \
  zotov@mi.ras.ru}}\end{center}

 \begin{abstract}
 We consider a family of classical elliptic integrable systems including
 (relativistic) tops and their matrix extensions of different types.
 These models can be obtained from the "off-shell" Lax pairs, which
 do not satisfy the Lax equations in general case but become true
 Lax pairs under various conditions (reductions).
 At the level of the off-shell Lax matrix there is a natural
 symmetry between the spectral parameter $z$ and relativistic parameter $\eta$. It is generated
 by the finite Fourier transformation, which we describe in detail.
 The symmetry allows to consider $z$ and $\eta$ on an equal
 footing. Depending on the type of integrable reduction any of the parameters can be chosen to be the spectral one. Then
 another one is the relativistic deformation parameter. As a
 by-product we describe the model of $N^2$ interacting $GL(M)$
 matrix tops and/or $M^2$ interacting $GL(N)$ matrix tops depending
 on a choice of the spectral parameter.
 \end{abstract}







\bigskip\bigskip\bigskip

\section{Integrable elliptic tops}
\setcounter{equation}{0}

We consider a special class of the Liouville integrable systems. The
simplest example is the (complexified) Euler top in ${\mathbb C}^3$:
 \beq\label{w01}
 \begin{array}{c}
  \displaystyle{
\dot {\vec S}={\vec S}\times J({\vec S})\,,
 }
 \end{array}
 \eq
 where ${\vec
S}=(S_1,S_2,S_3)$, $J({\vec S})=(J_1S_1,J_2S_2,J_3S_3)$ and
$\{J_i\}$ (inverse components of inertia tensor in principle axes)
are arbitrary constants. Equation (\ref{w01}) is simply rewritten in
matrix form:
 \beq\label{w02}
 \begin{array}{c}
  \displaystyle{
\dot S=[S,J(S)]\,,
 }
 \end{array}
 \eq
where
 \beq\label{w021}
 \begin{array}{c}
  \displaystyle{
S=\frac{1}{2\imath}\,\sum\limits_{k=1}^3\sigma_k S_k\,,\quad
J(S)=\frac{1}{2\imath}\,\sum\limits_{k=1}^3\sigma_k S_k J_k\,,
 }
 \end{array}
 \eq
and $\{\sigma_k\}$ are the Pauli matrices. Equation (\ref{w02})
admits generalization to $\Mat$-valued $S$. These model are known as
the Euler-Arnold tops \cite{Arnold,Arnold2,Mi1,Manakov,Mi2}. They
are not integrable for a generic linear constant operator $J$ but
only for very specific choices of $J$.

\noindent\paragraph{Elliptic non-relativistic top.} In particular,
the following model is integrable (and is called elliptic top)
\cite{STSR,LOZ,ZS}. Consider the Heisenberg group with generators
$Q$ and $\Lambda$: $\zeta Q\Lambda=\Lambda Q$. For
$\zeta=\exp(\frac{2\pi
 \imath}{N})$, $N\in\mZ_+$ it has
finite-dimensional representation given by $N\times
 N$ matrices
 \beq\label{w03}
 \begin{array}{c}
  \displaystyle{
Q_{kl}=\delta_{kl}\exp(\frac{2\pi
 \imath}{N}k)\,,\ \ \ \Lambda_{kl}=\delta_{k-l+1=0\,{\hbox{\tiny{mod}}}
 N}
 }
 \end{array}
 \eq
with the property $Q^N=\Lambda^N=1_{N\times N}$. Define the
following basis in $\Mat$ \cite{Belavin,RT}:
 \beq\label{w04}
 \begin{array}{c}
  \displaystyle{
 T_\al=T_{\al_1 \al_2}=\exp\left(\frac{\pi\imath}{N}\,\al_1
 \al_2\right)Q^{\al_1}\Lambda^{\al_2}\,,\quad
 \al=(\al_1,\al_2)\in\mZ_N\times\mZ_N\,.
 }
 \end{array}
 \eq
Then
  \beq\label{w05}
 \begin{array}{c}
  \displaystyle{
T_\al T_\be=\kappa_{\al,\be} T_{\al+\be}\,,\ \ \
\kappa_{\al,\be}=\exp\left(\frac{\pi \imath}{N}(\be_1
\al_2-\be_2\al_1)\right)\,,
 }
 \end{array}
 \eq
and
  \beq\label{w06}
 \begin{array}{c}
  \displaystyle{
[T_\al, T_\be]=C_{\al,\be} T_{\al+\be}\,,\ \ \
C_{\al,\be}=\kappa_{\al,\be}-\kappa_{\be,\al}\,,
 }
 \end{array}
 \eq
where we assume $\al+\be=(\al_1+\be_1,\al_2+\be_2)$. Using the basis
(\ref{w04}) we can define the inverse inertia tensor for the
elliptic top. It is given as follows:
 \beq\label{w07}
 \begin{array}{c}
  \displaystyle{
  S={\sum}'_\al T_\al S_\al\,,\quad J(S)={\sum}'_\al T_\al S_\al
J_\al\,,\quad J_\al=-\wp(\om_\al)\,,\quad
 \om_\al=\frac{\al_1+\al_2\tau}{N}\,,
 }
 \end{array}
 \eq
where the sums are over $\al\in\,\mZ_N\times\mZ_N$, the prime means
$\al\neq 0\equiv(0,0)$ and $\wp$ is the Weierstrass $\wp$-function
with elliptic moduli $\tau$ (Im$(\tau)>0$). In $N=2$ case this model
reproduces the Euler top (\ref{w021}). Indeed, in this case we have
three components of $S$ ($S_{(1,0)}$, $S_{(0,1)}$ and $S_{(1,1)}$).
The values of $J_\al$ are not arbitrary since they depend on a
single parameter $\tau$. However, one can shift them by the same
constant (without changing of equations of motion) and rescale
simultaneously by another constant (it is equivalent to rescaling of
time variable).

The integrability of (\ref{w02}) with $J$ (\ref{w07}) follows from
existence of the Lax pair with spectral parameter ($z$)
\cite{Skl1,STSR,LOZ}. It is similar to  Krichever's ansatz
\cite{Krich1} for elliptic Calogero-Moser model:
 \beq\label{w08}
 \begin{array}{c}
  \displaystyle{
  L(z)={\sum}'_\al T_\al S_\al\vf_\al(z,\om_\al)\,,\quad\quad
  M(z)={\sum}'_\al T_\al S_\al  f_\al(z,\om_\al)\,,
 }
 \end{array}
 \eq
 where the sets of function $\vf_\al(z,\om_\al)$ and
 $f_\al(z,\om_\al)$ ($\al\neq 0$) are defined by (\ref{w88}),
 (\ref{w90}).

The Lax equations
 \beq\label{w09}
 \begin{array}{c}
  \displaystyle{
  {\dot L}(z)=[L(z),M(z)]
 }
 \end{array}
 \eq
are equivalent to equations of motion of elliptic top (\ref{w02})
with $J$ (\ref{w07}) identically in $z$. The proof is based on
(\ref{w93}).

\paragraph{Relativistic elliptic top} \cite{LOZ8,LOZ81} is one-parameter deformation of
the elliptic top described above. Its Lax pair
 \beq\label{w10}
 \begin{array}{c}
  \displaystyle{
  L(z)={\sum}_\al T_\al S_\al\vf_\al(z,\eta+\om_\al)\,,\quad\quad
  M(z)=-{\sum}'_\al T_\al S_\al\vf_\al(z,\om_\al)
 }
 \end{array}
 \eq
provides equations of motion (\ref{w02}) with
 \beq\label{w11}
 \begin{array}{c}
  \displaystyle{
  J^\eta(S)={\sum}'_\al T_\al S_\al
J_\al^\eta\,,\quad\quad J_\al^\eta=E_1(\eta+\om_\al)-E_1(\om_\al)\,.
 }
 \end{array}
 \eq
Indeed, for the Lax pair (\ref{w10}) the l.h.s. of the Lax equations
(\ref{w09}) equals
 \beq\label{w12}
 \begin{array}{c}
  \displaystyle{
  {\dot L}(z)={\sum}_\al T_\al {\dot S}_\al\vf_\al(z,\eta+\om_\al)\,.
 }
 \end{array}
 \eq
For the r.h.s. of (\ref{w09}) we have:
 $$
 \begin{array}{c}
  \displaystyle{
 -{\sum}_\be{\sum}'_\ga [T_\be,T_\ga] S_\be S_\ga
 \vf_\be(z,\om_\be+\eta)\vf_\ga(z,\om_\ga)=-{\sum}'_{\be,\ga} [T_\be,T_\ga] S_\be S_\ga
 \vf_\be(z,\om_\be+\eta)\vf_\ga(z,\om_\ga)
 }
 \end{array}
 $$
since $T_{00}$ is identity matrix. (Anti)symmetrization with respect
to $\be$ and $\ga$ yields
 \beq\label{w13}
 \begin{array}{l}
  \displaystyle{
  -\frac12\,{\sum}'_{\be,\ga} [T_\be,T_\ga] S_\be S_\ga
  \Big(
  \vf_\be(z,\om_\be+\eta)\vf_\ga(z,\om_\ga)-\vf_\ga(z,\om_\ga+\eta)\vf_\be(z,\om_\be)\Big)\stackrel{(\ref{w06}),(\ref{w92})}{=}
 }
 \\ \ \\
   \displaystyle{
  -\frac12\,\,{\sum}'_{\be,\ga} C_{\be,\ga} T_{\be+\ga} S_\be S_\ga
  \vf_{\be+\ga}(z,\eta+\om_{\be+\ga})\Big( J_\be^\eta-J_\ga^\eta  \Big)=
 }
  \\ \ \\
   \displaystyle{
  {\sum}'_{\be,\ga} C_{\be,\ga} T_{\be+\ga} S_\be S_\ga
  \vf_{\be+\ga}(z,\eta+\om_{\be+\ga})J_\ga^\eta={\sum}'_{\al,\be} T_\al\vf_\al(z,\eta+\om_\al)
   C_{\be,\al-\be}
  S_\be S_{\al-\be} J_{\al-\be}^\eta \,.
 }
 \end{array}
 \eq
By equating coefficients behind $T_\al$ in (\ref{w12}) and
(\ref{w13}) we obtain
 \beq\label{w14}
 \begin{array}{c}
  \displaystyle{
  {\dot S}_{00}=0\,,\quad\quad {\dot S_\al}={\sum}'_{\be} C_{\be,\al-\be}
  S_\be S_{\al-\be} J_{\al-\be}^\eta\,,\quad \al\neq 0\,.
 }
 \end{array}
 \eq
Equations  of motion (\ref{w14}) are components of the matrix
equation (\ref{w02}) in the basis (\ref{w04})-(\ref{w06}) with
$J^\eta$  (\ref{w11}).

The non-relativistic limit (\ref{w07}) is obtained from (\ref{w14})
via rescaling of time variable by $\eta$, taking the limit
$\eta\rightarrow 0$ and usage of (\ref{w83}) and (\ref{w84}).

\paragraph{Relativization.} There is a simple relation between Lax
matrices of the non-relativistic (\ref{w08}) and relativistic
(\ref{w10}) tops. Denote the first one as $l(z,S)$ and the second
one as $L^\eta(z,S)$. Introduce also\footnote{Here and elsewhere we
mean $S_0=S_{00}$, and $1_N$ is the identity matrix of size
$N\!\times\! N$.} $L^0(z,S)=1_NS_0+l(z,S)$. Then
 \beq\label{w51}
 \begin{array}{c}
  \displaystyle{
  L^\eta(z-\eta,L^0(\eta,S))=\phi(z-\eta,\eta)L^0(z,S)\,.
 }
 \end{array}
 \eq
This follows from the definition of the Kronecker function
(\ref{w80}) because the latter implies
 \beq\label{w52}
 \begin{array}{c}
  \displaystyle{
  \frac{\vf_\al(z,\om_\al+\eta)}{\phi(z,\eta)}=\frac{\vf_\al(z+\eta,\om_\al)}{\vf_\al(\eta,\om_\al)}\,,\quad
\al\neq 0\,.
 }
 \end{array}
 \eq
Relation (\ref{w51}) means explicit change of variables
 \beq\label{w53}
 \begin{array}{c}
  \displaystyle{
S_\al\rightarrow S_\al/\vf_\al(\eta,\om_\al)
 }
 \end{array}
 \eq
between non-relativistic and relativistic models. At the same time
the transformation (\ref{w53}) changes eigenvalues of matrix $S$
which are obviously conserved by dynamics (\ref{w02}). In
particular, (\ref{w53}) can not connect non-relativistic and
relativistic
 models with rank 1 matrices $S$. The first one is known to be gauge
 equivalent to the elliptic Calogero-Moser model while the second is
 gauge equivalent to the Ruijsenaars-Schneider model.

\paragraph{$\mZ_2$ reduction.} The elliptic top (\ref{w07}) allows the following
reduction:
 \beq\label{w54}
 \begin{array}{c}
  \displaystyle{
S_\al=S_{-\al}\,,\quad\forall\al\,.
 }
 \end{array}
 \eq
At the level of the Lax matrix (\ref{w08}) it is generated by
condition
 \beq\label{w55}
 \begin{array}{c}
  \displaystyle{
 L(-z)=-hL(-z)h^{-1}\,,\quad h=\mathcal{J}\Lambda^{-1}\,,\quad
\mathcal{J}_{ij}=\delta_{i,N-j+1}\,.
 }
 \end{array}
 \eq
 Constraints (\ref{w54}) follows from (\ref{w55}) due to $h\,T_{\al}\,
 h^{-1}=T_{-\al}$. The general idea of $\mZ_2$ type reduction appeared first in \cite{Mi}.
 For details of (\ref{w54})-(\ref{w55}) see Appendix in \cite{LOZ16}.
 In relativistic case the set of constraints (\ref{w54}) is replaced
 by
 \beq\label{w56}
 \begin{array}{c}
  \displaystyle{
\frac{S_\al}{\vf_\al(\eta,\om_\al)}=\frac{S_{-\al}}{\vf_{-\al}(\eta,-\om_\al)}\,,\quad\forall\al
 }
 \end{array}
 \eq
 which is in agreement with the transition (\ref{w53}). The reduced
 models are integrable, and their Lax matrices are restrictions of
 the initial Lax matrices
 to the submanifolds in the phase spaces defined by (\ref{w54}) or (\ref{w56}) respectively.

This paper is a kind of a brief review of several previous articles
\cite{LOZ8,LOZ15,LOZ162,LOZ6,Z16}. The aim is have a fresh approach
and some new results by focus attention on possible Lax pairs and a
symmetry between relativistic and spectral parameters arizing from
the lattice Fourier transformation.

\section{Finite Fourier transformation on lattice $\mZ_N\times
\mZ_N$}\label{s2} \setcounter{equation}{0}
The following set of identities hold true:
 \beq\label{e913}
 \begin{array}{c}
  \displaystyle{
\frac{1}{N}\sum\limits_{\al} \kappa_{\al,\ga}^2\,
\vf_\al(N\hbar,\om_\al+\frac{z}{N})=\vf_\ga(z,\om_\ga+\hbar)
  }
 \end{array}
 \eq
 or, equivalently,
  \beq\label{e914}
 \begin{array}{c}
  \displaystyle{
\frac{1}{N}\sum\limits_{\al} \kappa_{\al,\ga}^2\,
\vf_\al(z,\om_\al+\hbar)=\vf_\ga(N\hbar,\om_\ga+\frac{z}{N})
  }
 \end{array}
 \eq
 with $\kappa_{\al,\ga}$ (\ref{w05}).
The proof can be found in \cite{LOZ15}. In the limit $z\rightarrow
0$ the identity (\ref{e914}) yields
  \beq\label{e915}
 \begin{array}{c}
  \displaystyle{
\frac{1}{N}\sum\limits_{\al}
\Big(E_1(\om_\al+\hbar)+2\pi\imath\,\p_\tau\om_\al\Big)=E_1(N\hbar)
  }
 \end{array}
 \eq
 for $\ga=0$ and
  \beq\label{e916}
 \begin{array}{c}
  \displaystyle{
\frac{1}{N}\sum\limits_{\al} \kappa_{\al,\ga}^2
\Big(E_1(\om_\al+\hbar)+2\pi\imath\,\p_\tau\om_\al\Big)=\vf_\ga(N\hbar,\om_\ga)
  }
 \end{array}
 \eq
 for $\ga\neq 0$. The simple pole in (\ref{e914}) at $z=0$ is cancelled due
 to\footnote{The permutation operator in the basis (\ref{w04}) is $P_{12}=(1/N)\sum_\al T_\al\otimes T_{-\al}$.
  In this respect (\ref{e9051}) is equivalent to $P_{12}^2=1_N\otimes 1_N$.}
  \beq\label{e9051}
 \begin{array}{c}
  \displaystyle{
\sum\limits_{\al} \ka_{\al,\ga}^2=N^2\,\delta_{\ga,0}.
 }
 \end{array}
 \eq
 Similarly, in the limit $\hbar\rightarrow 0$ the identity (\ref{e914}) yields
  \beq\label{e917}
 \begin{array}{c}
  \displaystyle{
\frac{1}{N}\,E_1(z)+\frac{1}{N}\sum\limits_{\al\neq 0}
\kappa_{\al,\ga}^2
\vf_\al(z,\om_\al)=E_1(\om_\ga+\frac{z}{N})+2\pi\imath\,\p_\tau\om_\ga\,.
  }
 \end{array}
 \eq
Taking the limit $\hbar\rightarrow 0$ in (\ref{e915}) or the limit
$z\rightarrow 0$ in (\ref{e917}) with $\ga=0$ we get:
  \beq\label{e918}
 \begin{array}{c}
  \displaystyle{
\frac{1}{N}\sum\limits_{\al\neq 0}
\Big(E_1(\om_\al)+2\pi\imath\,\p_\tau\om_\al\Big)=0\,,
  }
 \end{array}
 \eq
while the limit $\hbar\rightarrow 0$ in (\ref{e916}) or the limit
$z\rightarrow 0$ in (\ref{e917}) with $\ga\neq0$ yields:
  \beq\label{e919}
 \begin{array}{c}
  \displaystyle{
\frac{1}{N}\sum\limits_{\al\neq 0} \kappa_{\al,\ga}^2
\Big(E_1(\om_\al)+2\pi\imath\,\p_\tau\om_\al\Big)=E_1(\om_\ga)+2\pi\imath\,\p_\tau\om_\ga\,.
  }
 \end{array}
 \eq
The derivative of (\ref{e915}) with respect to $\hbar$ gives
  \beq\label{e920}
 \begin{array}{c}
  \displaystyle{
\sum\limits_{\al}
E_2(\om_\al+\hbar)=N^2E_2(N\hbar)\quad\hbox{or}\quad\sum\limits_{\al}
\wp(\om_\al+\hbar)=N^2\wp(N\hbar) \,.
  }
 \end{array}
 \eq
 Similarly from (\ref{e916}) for $\ga\neq 0$ we get
  \beq\label{e9202}
 \begin{array}{c}
  \displaystyle{
\sum\limits_{\al} \kappa_{\al,\ga}^2 E_2(\om_\al+\hbar)=-N^2
\vf_\ga(N\hbar,\om_\ga)
(E_1(N\hbar+\om_\ga)-E_1(N\hbar)+2\pi\imath\p_\tau\om_\ga)\,.
  }
 \end{array}
 \eq
With the limit $\hbar\rightarrow 0$ in (\ref{e920}) and the relation
(\ref{w84}) we obtain
  \beq\label{e921}
 \begin{array}{c}
  \displaystyle{
\sum\limits_{\al\neq 0}
\wp(\om_\al)=0\quad\hbox{or}\quad\sum\limits_{\al\neq 0}
E_2(\om_\al)=-\frac{N^2-1}{3}\,\frac{\vth'''(0)}{\vth'(0)}\,.
  }
 \end{array}
 \eq
The derivative of (\ref{e914}) with respect to $\hbar$ at $\hbar=0$
gives
  \beq\label{e922}
 \begin{array}{c}
  \displaystyle{
\frac{1}{2}\left(E_1^2(z)-\wp(z)\right)+\sum\limits_{\al\neq 0}
\kappa_{\al,\ga}^2 f_\al(z,\om_\al)\!=\!\frac{N^2}{2} \left(
(E_1(\om_\ga\!+\!\frac{z}{N})+2\pi\imath\p_\tau\om_\ga)^2-\wp(\om_\ga\!+\!\frac{z}{N})
\right)\,.

  }
 \end{array}
 \eq
Similarly, from (\ref{e913}) we have
  \beq\label{e923}
 \begin{array}{c}
  \displaystyle{
\sum\limits_{\al}\left(
(E_1(\om_\al\!+\!\frac{z}{N})+2\pi\imath\p_\tau\om_\al)^2-\wp(\om_\al\!+\!\frac{z}{N})\right)
=E_1^2(z)-\wp(z)
  }
 \end{array}
 \eq
and
  \beq\label{e924}
 \begin{array}{c}
  \displaystyle{
\frac{1}{2}\,\sum\limits_{\al} \kappa_{\al,\ga}^2 \left(
(E_1(\om_\al\!+\!\frac{z}{N})+2\pi\imath\p_\tau\om_\al)^2-\wp(\om_\al\!+\!\frac{z}{N})
\right)=f_\ga(z,\om_\ga)
  }
 \end{array}
 \eq
for $\ga\neq 0$.

 \section{Matrix extensions of tops}
\setcounter{equation}{0}

Let us apply the Fourier transformation (\ref{e913}) to the Lax
matrix of the relativistic top (\ref{w10}), where we replace $\eta$
by $\eta/N$:
 \beq\label{w61}
 \begin{array}{c}
  \displaystyle{
  L(z)=\sum\limits_\al T_\al
  S_\al\vf_\al(z,\om_\al+\frac{\eta}{N})=\sum\limits_\be\left(\frac{1}{N}\sum\limits_\al
  \kappa_{\be,\al}^2 T_\al
  S_\al\right) \vf_\be(\eta,\om_\be+\frac{z}{N})\,.
 }
 \end{array}
 \eq
The arguments $z$ and $\eta$ of functions $\vf_\ga$ are interchanged
but the coefficients behind $\vf_\be(\eta,\om_\be+\frac{z}{N})$ are
not proportional to $T_\be$, they are full matrices instead. It is
then natural to start with a more general matrix
 \beq\label{w62}
 \begin{array}{c}
  \displaystyle{
  A(z,\eta)=\sum\limits_\al
  A^\al\vf_\al(z,\om_\al+\frac{\eta}{N})\,,
 }
 \end{array}
 \eq
where $A^\al\in\hbox{Mat}(K,\mC)$. Then from (\ref{e913}) we
conclude that
 \beq\label{w63}
 \begin{array}{c}
  \displaystyle{
  A(z,\eta)=\sum\limits_\be {\ti A}^\be
  \vf_\be(\eta,\om_\be+\frac{z}{N})\,,
 }
 \end{array}
 \eq
where
 \beq\label{w64}
 \begin{array}{c}
  \displaystyle{
  {\ti A}^\be=\frac{1}{N}\sum\limits_\al
  \kappa_{\be,\al}^2 A^\al\,.
 }
 \end{array}
 \eq
In (\ref{w62})-(\ref{w64}) $N^2$ matrices $A^\al$ are of arbitrary
size $K$. Notice that if $K=N$ then the following two reductions to
the relativistic ${\rm gl}_N$ top are allowed:
 \beq\label{w641}
 \begin{array}{c}
  \displaystyle{
  1.\ A^\al=T_\al S_\al\quad\quad\quad 2.\ {\ti A}^\al=T_\al {\ti S}_\al
 }
 \end{array}
 \eq
In the first case $z$ is the spectral parameter while $\eta/N$ is
relativistic one. In the second case conversely $\eta$ is the
spectral parameter and $z/N$ is relativistic one. These two
reductions are of course incompatible.

In general case the matrix $A(z,\eta)$ is not Lax matrix of an
integrable system (or, at least, it is unknown) but there are
several more examples of integrable reductions besides (\ref{w64}).

\paragraph{Matrix tops.} If $K=NM$ then the following reductions are available (these are
the matrix extensions of the relativistic top \cite{LOZ16}):
 \beq\label{w65}
 \begin{array}{cc}
  1.\ A^\al=T_\al\otimes \bS_\al\,,\bS_\al\in\Matm\,,\quad\quad  & \quad\quad 2.\ {\ti A}^\al=T_\al\otimes
  \ti\bS_\al\,,\ti\bS_\al\in\Matm\,,
  \\ \ \\
 \bS_0=S_0 1_M \,,& \ti\bS_0=\ti S_0 1_M\,,
  \\ \ \\
  \displaystyle{
  \frac{\bS_\al}{\vf_\al(\eta/N,\om_\al)}=\frac{\bS_{-\al}}{\vf_{-\al}(\eta/N,-\om_\al)}
  }
  &
  \displaystyle{
  \frac{\ti\bS_\al}{\vf_\al(z/N,\om_\al)}=\frac{\ti\bS_{-\al}}{\vf_{-\al}(z/N,-\om_\al)}
  }
 \end{array}
 \eq
The left column case is described by the Lax pair
 \beq\label{w66}
 \begin{array}{c}
  \displaystyle{
{L^\eta}(z,\bS)=\sum\limits_{\al} T_\al\otimes \bS_\al\,
\vf_\al(z,\om_\al+\frac{\eta}{N})\,,\quad
M^\eta(z,\bS)=-\sum\limits_{\al\neq 0}T_\al\otimes
\bS_\al\,\vf_\al(z,\om_\al)\,,
 }
 \end{array}
 \eq
which provides equations of motion
 \beq\label{w67}
 \begin{array}{c}
  \displaystyle{
{\dot \bS}_\al=\sum\limits_{\be,\ga:\,\be+\ga=\al} \Big(
 \kappa_{\be,\ga}\,\bS_\be\bS_\ga - \kappa_{\ga,\be}\,\bS_\ga\bS_\be \Big)\,
 J_\ga^{\eta/N}\,,\ \al\neq 0;\ \
 J_\al^{\eta/N}=E_1(\frac{\eta}{N}+\om_\al)-E_1(\om_\al)\,.
 }
 \end{array}
 \eq
The constraints (the second and the third lines of the left column
of (\ref{w65})) are conserved on equations of motion. They are
sufficient conditions for the Lax equations to be fulfilled. When
$M=1$ the variables $\bS_\al$ become scalar $S_\al$ and (\ref{w67})
turns into (\ref{w14}).

In the Fourier dual case (right column of (\ref{w65})) the roles of
$z$ and $\eta$ are interchanged. This time $\eta$ is the spectral
parameter, and $z$ is the relativistic parameter. Equations of
motion and the Lax pair are obtained by replacement
$(z,\eta,\bS)\leftrightarrow(\eta,z,\ti\bS)$ in (\ref{w66}),
(\ref{w67}).

\paragraph{Gaudin like matrix tops on $\mZ_N\times \mZ_N$.} Let us return to the general case (\ref{w62}) with $N$
and $K$ be arbitrary integers. From (\ref{w63}) by changing
summation index $\be\rightarrow -\be$ we see that $A(z,\eta)$ as
matrix valued function on the elliptic curve with periods
$N,N\tau$
has simple poles at points
$N\om_\al=\al_1+\tau\al_2$ with residues $\ti A^{-\al}$. In this
respect it is the model of Gaudin type. Suppose the set of $N^2$
matrices $A^\al\in\hbox{Mat}(K,\mC)$, $\al\in\mZ_N\times \mZ_N$
satisfies
 \beq\label{w70}
 \begin{array}{c}
 \begin{array}{c}
1.\ A^0=A_0 1_K
\\ \ \\
\displaystyle{
  \frac{A^\al}{\vf_\al(\eta/N,\om_\al)}=\frac{A^{-\al}}{\vf_{-\al}(\eta/N,-\om_\al)}
  }
  \end{array}
 \quad\quad\quad\quad\quad
   \begin{array}{c}
2.\ \ti A^0=\ti A_0 1_K\,,
  \\ \ \\
  \displaystyle{
  \frac{\ti A^\al}{\vf_\al(z/N,\om_\al)}=\frac{\ti A^{-\al}}{\vf_{-\al}(z/N,-\om_\al)}
  }
  \end{array}
 \end{array}
 \eq
 Then (similarly to the previous case) the Lax equations with the Lax pair
 \beq\label{w71}
 \begin{array}{c}
  \displaystyle{
{L^\eta}(z,\{A^\al\})=\sum\limits_{\al} A^\al\,
\vf_\al(z,\om_\al+\frac{\eta}{N})\,,\quad
M^\eta(z,\{A^\al\})=-\sum\limits_{\al\neq
0}A^\al\,\vf_\al(z,\om_\al)\,,
 }
 \end{array}
 \eq
provides the following equations of motion
 \beq\label{w72}
 \begin{array}{c}
  \displaystyle{
{\dot A}^\al=\sum\limits_{\be,\ga:\,\be+\ga=\al} [A^\be,A^\ga]\,
 J_\ga^{\eta/N}\,,\ \al\neq 0;\ \
 J_\al^{\eta/N}=E_1(\frac{\eta}{N}+\om_\al)-E_1(\om_\al)\,.
 }
 \end{array}
 \eq
Again, the Lax pair and equations of motion for the Fourier dual
model are obtained by replacement
$(z,\eta,A^\al)\leftrightarrow(\eta,z,\ti A^\al)$, i.e. the Lax
matrices are of the same form due to
${L^\eta}(z,\{A^\al\})={L^z}(\eta,\{\ti A^\al\})=A(z,\eta)$ but with
different reductions constraints, and the $M$-matrices are
different.

\paragraph{(In)compatibility of reductions and Euler case.}
In general case any reduction of the first type is incompatible with
a reduction of the second type. It happens because (\ref{w641})
leaves only one non-zero matrix component for each matrix $A^\al$
(or $\ti A^\al$), while the reductions (\ref{w65}) and (\ref{w70})
explicitly depend on the corresponding relativistic parameter. The
latter plays the role of the spectral parameter in Fourier dual type
of reduction but dynamical variables should be independent of the
spectral parameter. However, there is a special case when two
reductions of different types are compatible. It is $N=2$ case
because the second line of (\ref{w70}) (or the third line of
(\ref{w65})) is fulfilled by itself due to
$\al=-\al\,$mod$\,\mZ_2\times \mZ_2$ in this case. Therefore, the
conditions given by the second line of (\ref{w70}) (or the third
line of (\ref{w65})) are identities. This case is the matrix
extension of the Euler top (\ref{w01})-(\ref{w021}). Consider two
reductions in (\ref{w70}). The matrix of the Fourier transformation
$\kappa^2_{\al,\be}/N$ is of the size $N^2\times N^2$ because
$\al,\be\in\mZ_N\times \mZ_N$. In $N=2$ case it is $4\times 4$:
 \beq\label{w79}
 \begin{array}{c}
 \left(\begin{array}{c}
 \ti A^{(0,0)}
 \\
  \ti A^{(1,0)}
 \\
  \ti A^{(0,1)}
 \\
  \ti A^{(1,1)}
 \end{array}\right)
 =
\displaystyle{ \frac12}\,\left(
  \begin{array}{cccc}
  1 & 1 & 1 & 1
  \\
  1 & 1 & -1 & -1
  \\
    1 & -1 & 1 & -1
  \\
    1 & -1 & -1 & 1
   \end{array}
   \right)
    \left(\begin{array}{c}
  A^{(0,0)}
 \\
   A^{(1,0)}
 \\
   A^{(0,1)}
 \\
   A^{(1,1)}
 \end{array}\right)
 \end{array}
 \eq
 If all conditions of (\ref{w70}) are satisfied then
 \beq\label{w791}
 \begin{array}{c}
A^{(0,0)}\sim 1_K\ \ \hbox{and}\ \ \ti
A^{(0,0)}=\frac12(A^{(0,0)}+A^{(0,1)}+A^{(1,0)}+A^{(1,1)})\sim
1_K\,,
 \end{array}
 \eq
and we are left with three matrices $A^{(0,1)},A^{(1,0)},A^{(1,1)}$
satisfying $A^{(0,1)}+A^{(1,0)}+A^{(1,1)}\sim 1_K$.

\section{${\rm GL}_N\times{\rm GL}_M$ models}
\setcounter{equation}{0}

In this section we describe ${\rm GL}_N\times{\rm GL}_M$ models.
They are generalizations of the relativistic top
(\ref{w10})-(\ref{w11}), where dependence on the parameter $\eta$ is
similar to dependence on the spectral parameter $z$. The study of
such models is inspired by $R$-matrix description of integrable
tops, and existence of the symmetric $R$-matrix with (almost)
symmetric dependence on the spectral parameter and the Planck
constant. We review briefly these constructions, and then describe
${\rm GL}_N\times{\rm GL}_M$ models in much the same way as in the
previous section. As a by-product of the $R$-matrix construction we
have simple formulation (\ref{w1005}) of the Fourier transformations
(\ref{e913}).
\paragraph{$R$-matrix description.} The Lax pairs, equations of
motion and Hamiltonian description of integrable tops under
consideration can be formulated in terms of $GL(N)$ quantum
$R$-matrix $R_{12}^\hbar(z)\in\Mat^{\otimes 2}$ satisfying the
following properties:
 \beq\label{w1001}
 \begin{array}{l}
  \displaystyle{
\hbox{unitarity:}\quad R_{12}^\hbar(z)R_{21}^\hbar(-z)\sim
1_N\otimes 1_N\,,
   }
   \\ \ \\
  \displaystyle{ \hbox{associative Yang-Baxter equation \cite{FK,Pol}:}}\\ \ \\
    \displaystyle{
 R^\hbar_{12}(z_1-z_2)
 R^{\eta}_{23}(z_2-z_3)=R^{\eta}_{13}(z_1-z_3)R_{12}^{\hbar-\eta}(z_1-z_2)+R^{\eta-\hbar}_{23}(z_2-z_3)R^\hbar_{13}(z_1-z_3)\,,
 }
     \\ \ \\
  \displaystyle{
\hbox{classical limit (expansion near } \hbar=0):
  }
       \\ \ \\
  \displaystyle{
R_{12}^\hbar(z)=\frac{1}{\hbar}\,1\otimes 1
+r_{12}(z)+\hbar\,m_{12}(z)+O(\hbar^2)\,.
  }
 \end{array}
 \eq
Namely, it can be shown that with the properties (\ref{w1001}) the
Lax pair
 \beq\label{w1002}
 \begin{array}{c}
  \displaystyle{
L^\eta(z,S)=\tr_2(R^\eta_{12}(z)S_2)\,, \quad
M^\eta(z,S)=-\tr_2(r_{12}(z){ S}_2)
  }
 \end{array}
 \eq
provides equations of motion (\ref{w02})
 \beq\label{w1003}
 \begin{array}{c}
  \displaystyle{
{\dot L}^\eta(z,S)=[L^\eta(z,S),M^\eta(z,S)]\,,\quad \forall z\quad
\Longleftrightarrow\quad \dot S=[S,J^\eta(S)]\,.
  }
 \end{array}
 \eq
  In the elliptic case the
 $R$-matrix is the Belavin's one \cite{Belavin}. Being written in the
 form
 \beq\label{w1004}
 \begin{array}{c}
  \displaystyle{
R_{12}^\hbar(z)=\sum\limits_\al T_\al\otimes T_{-\al}
\vf_\al(z,\om_\al+\hbar)
  }
 \end{array}
 \eq
it reproduces\footnote{The $M$-matrix from (\ref{w1002}) differs
from the one (\ref{w10}) by scalar component $1_NE_1(z)S_{(0,0)}$
which is cancelled out from the Lax equations.} the Lax pair
(\ref{w10}) via (\ref{w1002}). Details of this construction can be
found in \cite{LOZ16}. It is based on the fact that $R$-matrix
(\ref{w1004}) with properties (\ref{w1001}) can be viewed as matrix
generalization of the Kronecker function (\ref{w80})
\cite{FK,LOZ14,LOZ15,Z16}. In particular, the trivial property
$\phi(z,\hbar)=\phi(\hbar,z)$ of the Kronecker function is
generalized to
 \beq\label{w1005}
 \begin{array}{c}
  \displaystyle{
R_{12}^\hbar(z)P_{12}=R_{12}^{z/N}(N\hbar)\,,\quad\quad
P_{12}=\frac1N\sum\limits_\al T_\al\otimes T_{-\al}\,,
  }
 \end{array}
 \eq
 where $P_{12}$ is the permutation operator. The latter relation
 (\ref{w1005}) is equivalent to the set of finite Fourier
 transformations (\ref{e913}).
\paragraph{Symmetric $R$-matrix.} Functions $\vf_\al(z,\om_\al+\eta)$ entering $R$-matrix
(\ref{w1004}) as well as the Lax matrix of the relativistic top
(\ref{w10}) and its matrix extension (\ref{w62}) have a single
simple pole at $z=0$ on the elliptic curve $\Sigma_{z,\tau}$ with
coordinate $z$ and moduli $\tau$ (and periods $1,\tau$). On the
other hand, as functions of $\eta$ they have simple poles at
$\eta=-\om_\al$. In $GL(N)\times GL(M)$ case there is a possibility
to make dependence on $z$ and $\eta$ in much the same way.

Let $N$ and $M$ be positive integers. For simplicity we also assume
that $N$ and $M$ are coprime g.c.d.$(N,M)=1$. Consider the following
set of $N^2M^2$ functions
 \beq\label{w15}
 \begin{array}{c}
  \displaystyle{
   \Phi_{\al,\ti\al}(z,\eta)=
  \exp\left( 2\pi\imath(z+N{\ti\om}_{\ti\al})\frac{\al_2}{N} + 2\pi\imath\eta\, N\frac{{\ti\al}_2}{M} \right)
   \phi(z+N{\ti\om}_{\ti\al}\,,\eta+\om_\al)=
   }
   \\ \ \\
  \displaystyle{
=\exp\left( 2\pi\imath\eta\, N\frac{{\ti\al}_2}{M} \right)
   \vf_\al(z+N{\ti\om}_{\ti\al}\,,\eta+\om_\al)\,,
  }
 \end{array}
 \eq
 where
 $$
 \al=(\al_1,\al_2)\in\mZ_N\times\mZ_N\,,\
 \om_\al=\frac{\al_1+\al_2\tau}{N}\,,\
 \tal=(\tal_1,\tal_2)\in\mZ_M\times\mZ_M\,,\
 \tom_\tal=\frac{\tal_1+\tal_2\tau}{M}\,.
 $$
We will keep tildes for the variables related to $\Matm$. The
functions $\Phi_{\al,\ti\al}(z,\eta)$ are almost symmetrical (but
not completely) with respect to simultaneous interchanging of
$z\leftrightarrow\eta$ and $N\leftrightarrow M$. The reason for
non-symmetric dependence comes from our additional requirement on
$\Phi_{\al,\ti\al}(z,\eta)$ to be periodic with respect to shifts of
discrete variables:
 \beq\label{w16}
 \begin{array}{c}
  \displaystyle{
   \Phi_{(\al_1+N,\al_2),\ti\al}(z,\eta)=\Phi_{(\al_1,\al_2+N),\ti\al}(z,\eta)=\Phi_{\al,\ti\al}(z,\eta)\,,
   }
   \\ \ \\
  \displaystyle{
   \Phi_{\al,(\tal_1+M,\tal_2)}(z,\eta)=\Phi_{\al,(\tal_1,\tal_2+M)}(z,\eta)=\Phi_{\al,\tal}(z,\eta)\,.
  }
 \end{array}
 \eq
The properties (\ref{w16}) can be verified using (\ref{w841}). With
these properties the functions $\Phi_{\al,\ti\al}(z,\eta)$ are well
defined on $\al\in \mZ^2$, $\tal\in \mZ^2$ -- they are periodic with
periods $N$ and $M$. The latter is equivalent to the definition of
the functions on $\mZ_N^2\times \mZ_M^2$ and convenient for
summation (subtraction) of indices.

Analogues of identities (\ref{w91}) and (\ref{w92}) are easily
obtained. Due to (\ref{w15})
$\Phi_{\al,\tal}(z,0)=\vf_\al(z+N{\ti\om}_{\ti\al}\,,\om_\al)$, and
we get the following identity from (\ref{w91}):
 \beq\label{w33}
 \begin{array}{c}
  \displaystyle{
\Phi_{\be,\tbe}(z,\eta)\,\Phi_{\ga,\tga}(z,0)=
 \Phi_{\be,\tbe-\tga}(0,\eta)\,\Phi_{\be+\ga,\tga}(z,\eta)+\Phi_{\ga,\tga-\tbe}(0,0)\,\Phi_{\be+\ga,\tbe}(z,\eta)\,.
  }
 \end{array}
 \eq
It is valid for $\ga\neq 0$ and $\tbe\neq\tga$. If $\tbe=\tga$ then
from (\ref{w92}) we have
 \beq\label{w34}
 \begin{array}{l}
  \displaystyle{
\Phi_{\be,\tbe}(z,\eta)\,\Phi_{\ga,\tbe}(z,0)=
 }
 \\ \ \\
  \displaystyle{
=\Phi_{\be+\ga,\tbe}(z,\eta)
 (E_1(z+N\tom_\tbe)+E_1(\eta+\om_\be)+E_1(\om_\ga)-E_1(z+\eta+\om_{\be+\ga}+N\tom_\tbe))
\,.
  }
 \end{array}
 \eq
Alternatively,
 \beq\label{w331}
 \begin{array}{c}
  \displaystyle{
\Phi_{\be,\tbe}(z,\eta)\,\Phi_{\ga,\tga}(0,\eta)=
 \Phi_{\be-\ga,\tbe}(z,0)\,\Phi_{\ga,\tbe+\tga}(z,\eta)+\Phi_{\ga-\be,\tga}(0,0)\,\Phi_{\be,\tbe+\tga}(z,\eta)\,,
  }
 \end{array}
 \eq
which is valid for $\tga\neq 0$ and $\be\neq\ga$. In other cases
 \beq\label{w341}
 \begin{array}{l}
  \displaystyle{
\Phi_{\be,\tbe}(z,\eta)\,\Phi_{\be,\tga}(0,\eta)=
 }
 \\ \ \\
  \displaystyle{
=\Phi_{\be,\tbe+\tga}(z,\eta)
 (E_1(z+N\tom_\tbe)+E_1(N\tom_\tga)+E_1(\eta+\om_\be)-E_1(z+\eta+N\tom_{\tbe+\tga}+\om_\be))
\,.
  }
 \end{array}
 \eq

The set of functions (\ref{w15}) allows to define the so-called
symmetric $R$-matrix \cite{LOZ162}:
 \beq\label{w17}
 \begin{array}{c}
  \displaystyle{
   \mR_{12,\ti 1\ti 2}(z,\hbar)
=\sum\limits_{\al\in\, {\mathbb Z}_N\times {\mathbb Z}_N}
   \sum\limits_{\ti\al\in\, {\mathbb Z}_M\times {\mathbb
   Z}_M} \Phi_{\al,\ti\al}(z,\hbar)\,T_\al\otimes T_{-\al}\otimes{\ti T}_{\ti\al}\otimes {\ti
   T}_{-\ti\al}\,,
   }
 \end{array}
 \eq
 where $\{\tT_\tal\}$ -- is the basis in $\Matm$
(\ref{w04})-(\ref{w06}) with $N$ replaced by $M$. It satisfies the
unitarity condition $\mR_{12,\ti 1\ti 2}\,\mR_{21,\ti 1\ti
2}=1_N\otimes 1_N\otimes {\ti 1}_M\otimes {\ti
1}_M\,N^2M^2(\wp(N\hbar)-\wp(Mz))$ and the associative Yang-Baxter
equation
  \beq\label{w18}
  \begin{array}{c}
  \displaystyle{
\mR_{12,\ti 1\ti 2}\, \mR_{23,\ti 3\ti 2}=\mR_{13,\ti 3\ti 2}\,
\mR_{12,\ti 1\ti 3} + \mR_{23,\ti 3\ti 1}\,\mR_{13,\ti 1\ti 2}\,,
 }
 \end{array}
 \eq
  where $\mR_{ab,\ti a\ti b}=\mR_{ab,\ti a\ti b}(z_{a}-z_b,\hbar_{\ti a}-\hbar_{\ti
  b})$. These are the first and the second conditions from
  (\ref{w1001}). The third condition (classical limit) is not
  fulfilled because the simplest rational analogue of (\ref{w17}) is
  \beq\label{w19}
  \begin{array}{c}
  \displaystyle{
 \mR_{12,\ti 1\ti 2}(z,\hbar)=M\frac{1_{N}\otimes
1_{N}\otimes {\ti P}_{\ti 1\ti 2}}{\hbar}+N\frac{P_{12}\otimes {\ti
1}_M\otimes {\ti 1}_M}{z}\,.
 }
 \end{array}
 \eq
 So, the expansion of $\mR_{12,\ti 1\ti 2}(z,\hbar)$ (\ref{w17})
 near $\hbar=0$ does not starts with $1_{NM}\otimes 1_{NM}$ as well
 as similar expansion near $z=0$. At the same time in $M=1$ case
 (\ref{w17}) reproduces $R_{12}^\hbar(z)$ (\ref{w1004}), and for
 $N=1$ it yields $R^z_{12}(\hbar)$:
%
  \beq\label{w191}
  \begin{array}{ccc}
    &  \ \mR_{12,\ti 1\ti 2}(z,\hbar) &
 \\
   ^{M=1} \swarrow& &   \searrow^{N=1}
 \\
 R_{12}^{\,\hbar}(z) &    & R_{\ti 1 \ti 2}^{\,z}(\hbar)
 \end{array}
 \eq
 In this respect $\mR_{12,\ti 1\ti
 2}(z,\hbar)$ is an intermediate case between $GL(NM)$-valued
 Belavin's $R$-matrices $R^z_{12}(\hbar)$ and $R_{12}^\hbar(z)$.
 More precisely, since $\mZ_{NM}\cong\mZ_N\times\mZ_M$ ($N$ and $M$ are
 coprime) we can perform the Fourier transformation on sublattice $\mZ_N\times
 \mZ_N$ or $\mZ_M\times\mZ_M$:
  \beq\label{w20}
  \begin{array}{c}
  \displaystyle{
 \mR_{12,\ti 1\ti 2}(z,\hbar)\,P_{12}\otimes {\ti
1}_M\otimes {\ti 1}_M=R_{12}^{z/N}(N\hbar) }\,,
\\ \ \\
  \displaystyle{
 \mR_{12,\ti 1\ti 2}(z,\hbar)\,1_{N}\otimes
1_{N}\otimes {\ti P}_{\ti 1\ti 2}=R_{12}^{\hbar/M}(Mz)}\,,
 \end{array}
 \eq
where in r.h.s. we mean $R$-matrix (\ref{w1004}) with $N:=NM$.
\paragraph{${\rm GL}_N\times{\rm GL}_M$ models.} The
fact that the classical limit of (\ref{w17}) is not of the form
required in (\ref{w1001}) (in both -- $z$ and $\hbar$) means that
the ansatz (\ref{w1002}) does not work\footnote{Equivalently, the
pair of matrices $L^\eta(z)=\sum_{\al,\ti\al} S_{\al,\ti\al}
T_\al\otimes \ti T_{\ti\al} \Phi_{\al,\ti\al}(z,\eta)$ and
$M^\eta(z)=-\sum_{\al\neq 0,\ti\al}  S_{\al,\ti\al} T_\al\otimes \ti
T_{\ti\al} \Phi_{\al,\ti\al}(z,\eta)$ does not satisfy Lax equations
identically in $z$ or $\eta$.}. However, it is not of great
importance for matrix extensions because of additional constraints.
For example, the Lax pair of the matrix top (\ref{w66}) is related
through (\ref{w1002}) with $R$-matrix $R^\eta_{12,\ti 1\ti
2}(z)=R^\eta_{12}(z)\otimes {\ti P}_{\ti 1 \ti 2}$, which has the
wrong first term in the classical limit. This problem is solved by
imposing constraints (\ref{w65}). The same phenomenon takes place
for $\mR_{12,\ti 1\ti 2}(z,\hbar)$.

%
Consider the following matrix-valued function on
$\Sigma_{z,\tau}\times \Sigma_{\eta,\tau}$:
 \beq\label{w30}
 \begin{array}{c}
  \displaystyle{
  {\mathcal A}(z,\eta)=\sum_{\al,\tal} {\mathcal A}^{\al,\tal}\, \Phi_{\al,\tal}(z,\eta)\in {\rm Mat}(K,\mathbb
  C)\,,\quad \al\in\,\mZ_{N}^{\times 2}\,,\ \ti\al\in\,\mZ_{M}^{\times 2}\,.
  }
 \end{array}
 \eq
Suppose first that the spectral parameter is $z$. Using the
definitions (\ref{w15}) and (\ref{w88}) rewrite
$\Phi_{\al,\tal}(z,\eta)$ as follows:
 \beq\label{w301}
 \begin{array}{c}
  \displaystyle{
  \Phi_{\al,\tal}(z,\eta)=\ti\ka_{\al,\ti\al}^2\exp(2\pi\imath
  z\frac{\al_2}{N})\,
  \vf_{N\ti\al}(\eta+\om_\al,z+N\ti\om_{\ti\al})\,,\quad
  \ti\ka_{\al,\ti\al}=\exp(\pi\imath\frac{\ti\al_1\al_2-\al_1\ti\al_2}{M})\,.
  }
 \end{array}
 \eq
By applying the Fourier transform on $\mZ_M\times\mZ_M$ (\ref{e914})
 we get
 \beq\label{w302}
 \begin{array}{c}
  \displaystyle{
  \Phi_{\al,\tal}(z,\eta)=\frac{1}{M}\sum\limits_{\ti\ga}\ti\ka_{\al+N\ti\ga,\ti\al}^2\,
  \vf_{\al+N\ti\ga}(Mz,\ti\om_{\ti\ga}+\frac{\om_\al}{M}+\frac{\eta}{M})\,,
  }
 \end{array}
 \eq
where
$\ti\om_{\ti\ga}+\frac{\om_\al}{M}=\frac{1}{M}\,\om_{\al+N\ti\ga}\in\mZ_{NM}\oplus\tau\mZ_{NM}$.
Therefore, denoting $\al+N\ti\ga=a\in\mZ_{NM}^{\times 2}$ we come to
 \beq\label{w303}
 \begin{array}{c}
  \displaystyle{
  {\mathcal A}(z,\eta)=\sum\limits_{a\in\,\mZ_{NM}^{\times 2}} {\ti
  {\mathcal A}}^a\vf_a(Mz,\frac{\om_a+\eta}{M})\,,\quad {\ti {\mathcal A}}^a={\ti {\mathcal A}}^{\al+N\ti\ga}=
  \frac{1}{M}\sum\limits_{\ti\al\in\,\mZ_{M}^{\times 2}}\ti\ka_{\al+N\ti\ga,\ti\al}^2\,
  {\mathcal A}^{\al,\ti\al}\,,
  }
 \end{array}
 \eq
i.e. to the matrix function of form (\ref{w62}) with $N:=NM$, which
become Lax matrix in cases (\ref{w641}), (\ref{w65}) and
(\ref{w70}).

Suppose now that the spectral parameter in (\ref{w30}) is $\eta$.
Similarly to the previous case, apply the Fourier transform
(\ref{e913}) to $\Phi_{\al,\tal}(z,\eta)$ on the lattice
$\mZ_N\times \mZ_N$:
 \beq\label{w304}
 \begin{array}{c}
  \displaystyle{
  \Phi_{\al,\tal}(z,\eta)=\frac{1}{N}\exp(2\pi\imath N\eta\frac{\ti\al_2}{M})
  \sum\limits_{\ga}\ka^2_{\ga,\al}\vf_{\ga}(N\eta,\ti\om_{\ti\al}+\om_\ga+\frac{z}{N})\,.
  }
 \end{array}
 \eq
Again\footnote{Here we use that g.c.d.$(N,M)=1$}
$\om_{\ti\al}+\om_\ga=\frac{1}{N}\,\ti\om_{N\ti\al+M\ga}\in\mZ_{NM}\oplus\tau\mZ_{NM}$.
Therefore, denoting $M\ga+N\ti\al=a\in \mZ_{NM}^{\times 2}$ we have
 \beq\label{w305}
 \begin{array}{c}
  \displaystyle{
  {\mathcal A}(z,\eta)=\sum\limits_{a\in\,\mZ_{NM}^{\times 2}} {\ti
  {\mathcal A}}^a\vf_a(N\eta,\frac{\om_a+z}{N})\,,\quad {\ti {\mathcal A}}^a={\ti {\mathcal A}}^{N\ti\al+M\ga}=
  \frac{1}{N}\sum\limits_{\al\in\,\mZ_{N}^{\times 2}}\ka_{\ga,\al}^2\,
  {\mathcal A}^{\al,\ti\al}\,,
  }
 \end{array}
 \eq
i.e. the matrix function of the form (\ref{w62}) with $N:=NM$.

Conversely, we can start with $A(x,y)$ in the form (\ref{w62}):
 \beq\label{w306}
 \begin{array}{c}
  \displaystyle{
  A(x,y)=
  \sum\limits_{a\in \mZ_{NM}^{\times 2}}
  A^a\vf_a(x,\om_a+y)=
 }
 \end{array}
 \eq
 $$
=\sum\limits_{\al\in \mZ_{N}^{\times 2}}\sum\limits_{\ti\al\in
\mZ_{M}^{\times 2}}
  A^{\al,\ti\al}\exp(2\pi\imath
  x\,\frac{M\al_2+N\ti\al_2}{NM})\,\phi(x,\om_\al+\ti\om_{\ti\al}+y)\,,
 $$
 and apply the Fourier transformations on $\mZ_{N}^{\times 2}$ and $\mZ_{M}^{\times
 2}$:
  \beq\label{w3061}
 \begin{array}{c}
  \displaystyle{
  \exp(2\pi\imath
  x\,\frac{M\al_2+N\ti\al_2}{NM})\,\phi(x,\om_\al+\ti\om_{\ti\al}+y)=
  \frac{1}{N}\sum\limits_{\ga\in\,\mZ_{N}^{\times
  2}}\kappa_{\ga,\al}^2 \Phi_{\ga,\ti\al}(Ny,\frac{x}{N})\,,
 }
 \end{array}
 \eq
  \beq\label{w3062}
 \begin{array}{c}
  \displaystyle{
  \exp(2\pi\imath
  x\,\frac{M\al_2+N\ti\al_2}{NM})\,\phi(x,\om_\al+\ti\om_{\ti\al}+y)=
  \frac{1}{M}\sum\limits_{\tga\in\,\mZ_{M}^{\times
  2}}\ti\kappa_{\ti\ga,\ti\al}^2
  \ti\Phi_{\ti\ga,\al}(My,\frac{x}{M})\,,
 }
 \end{array}
 \eq
 where $\ti\Phi$ is obtained from $\Phi$ (\ref{w15}) by interchanging
 $N$ and $M$.

First, perform the Fourier transform on $\mZ_{N}^{\times 2}$
(\ref{w3061}). Then
 \beq\label{w307}
 \begin{array}{c}
  \displaystyle{
  A(x,y)=\sum\limits_{\ga\in\mZ_{N}^{\times 2}}\sum\limits_{\ti\al\in
\mZ_{M}^{\times 2}} {\ti A}^{\ga,\ti\al}
\Phi_{\ga,\ti\al}(Ny,\frac{x}{N})\,,\quad {\ti
A}^{\ga,\ti\al}=\frac{1}{N} \sum\limits_{\al\in \mZ_{N}^{\times 2}}
\ka^2_{\ga,\al} A^{\al,\ti\al}\,.
 }
 \end{array}
 \eq
Second, perform the Fourier transform on $\mZ_{M}^{\times 2}$
(\ref{w3062}). Then
 \beq\label{w308}
 \begin{array}{c}
  \displaystyle{
  A(x,y)=\sum\limits_{\al\in\mZ_{N}^{\times 2}}\sum\limits_{\ti\ga\in
\mZ_{M}^{\times 2}} {\ti A}^{\al,\ti\ga}
\ti\Phi_{\ti\ga,\al}(My,\frac{x}{M})\,,\quad {\ti
A}^{\al,\ti\ga}=\frac{1}{M} \sum\limits_{\ti\al\in \mZ_{M}^{\times
2}} {\ti\ka}^2_{\ti\ga,\ti\al} A^{\al,\ti\al}\,.
 }
 \end{array}
 \eq
In both cases (\ref{w307}) and (\ref{w308}) we can substitute $x=z$,
$y=\eta$ or $x=\eta$, $y=z$. These two substitutions are related by
the Fourier transform (\ref{w63}), (\ref{w64}) on $\mZ_{NM}^{\times
2}$. Thus, there are four natural possible forms of the matrix
function (\ref{w62}), and (\ref{w30}) is one of these.
\paragraph{Matrix Gaudin models.} The matrix (\ref{w30}) can be viewed as matrix generalization
of the special Gaudin models with marked points at $\{\om_\al\}$ or
$\{\ti\om_{\ti\al}\}$. Indeed, the Gaudin model based on the
relativistic top (\ref{w10}) is defined by the Lax
matrix\footnote{This model is the quasi-classical limit of
generalized spin chain with inhomogeneous parameters $z_k$ and
$L(z)$ being the monodromy matrix. In quantum case $L(z)$ satisfies
exchange relations which provides quadratic algebras of Sklyanin's
type \cite{Skl1,LOZ6}.}
 \beq\label{w309}
 \begin{array}{c}
  \displaystyle{
L(z)=\sum\limits_{k=1}^n\sum\limits_\al T_\al
S^k_\al\vf_\al(z-z_k,\eta+\om_\al)\,.
 }
 \end{array}
 \eq
The set of matrices $S^k={\sum}_\al T_\al S^k_\al$ are residues at
$z=z_k$. In our case the marked points $z_k$ are given by the set
$\{\om_\al\,,\al\in\mZ_N^{\times2}\}$ or
$\{\ti\om_{\ti\al}\,,\al\in\mZ_M^{\times2}\}$. To be precise,
consider (\ref{w307}) with $y=z/N$ and $x=N\eta$ ($z$ is supposed to
be spectral parameter). Impose reduction
 \beq\label{w310}
 \begin{array}{c}
  \displaystyle{
{\ti A}^{\ga,\ti\al}=\exp(-2\pi\imath
\eta{\ti\al_2})S^{-N\ti\al}_\ga\, T_\ga\,.
 }
 \end{array}
 \eq
Then the matrix (\ref{w307}) acquires the form (\ref{w309}) with
marked points at $z=-N\ti\om_{\ti\al}$. Similarly, consider
(\ref{w308}) with $y=z/M$ and $x=M\eta$  ($z$ is again the spectral
parameter) and impose reduction
 \beq\label{w311}
 \begin{array}{c}
  \displaystyle{
{\ti A}^{\al,\ti\ga}=\exp(-2\pi\imath
\eta{\al_2})S^{-M\al}_{\ti\ga}\, {\ti T}_{\ti\ga}\,.
 }
 \end{array}
 \eq
Then the matrix (\ref{w307}) acquires the form (\ref{w309}) with
marked points at $z=-M\om_{\al}$. The same reductions can be made
for the case when $\eta$ is treated as spectral parameter.
Thus, the off-shell (unreduced) matrix (\ref{w306}) contains both -
matrix extension of $GL(M)$ Gaudin model with $N^2$ marked points at
$\om_\al$ and $GL(N)$ Gaudin model with $M^2$ marked points at
$\ti\om_{\ti\al}$.\footnote{In transition between these two cases
one should rescale $z\leftrightarrow Nz/M$ and $\eta\leftrightarrow
M\eta/N$.} Moreover, in both cases any one of the parameters ($z$ or
$\eta$) can be considered as the spectral.

 \section{Conclusion and discussion}

In this paper we consider a wide class of integrable relativistic
elliptic tops and their different matrix extensions. The simplest
model
 \beq\label{w141}
 \begin{array}{c}
  \displaystyle{
  {\dot S}_{00}=0\,,\quad\quad {\dot S_\al}={\sum}'_{\be} C_{\be,\al-\be}
  S_\be S_{\al-\be} J_{\al-\be}^\eta\,,\quad \al\neq 0\,,\quad J_\al^\eta=E_1(\eta+\om_\al)-E_1(\om_\al)
 }
 \end{array}
 \eq
is a generalization of the Euler-Arnold top for multi-dimensional
(complexified) space. For rank one matrix $S$ it is gauge equivalent
to the elliptic Ruijsenaars-Schneider model with relativistic
parameter $\eta$.  Contrary to the many-body systems the set of
functions defining the underlying bundle (the Lax matrix is its
section) is independent of dynamical variables. This allows to treat
the spectral and relativistic parameters on equal terms.

First we review results of \cite{LOZ8} and \cite{LOZ16}. The main
point of that constructions is that underlying $R$-matrices should
satisfy the associative Yang-Baxter equation (\ref{w1001}) together
with some additional requirements such as skew-symmetry (or
unitarity) and the classical limit should begin with the identity
matrix. The latter condition can be violated. For example, equations
(\ref{w67}) or (\ref{w72})
 \beq\label{w721}
 \begin{array}{c}
  \displaystyle{
{\dot A}^\al=\sum\limits_{\be,\ga:\,\be+\ga=\al} [A^\be,A^\ga]\,
 J_\ga^{\eta/N}\,,\ \al\neq 0;\ \
 J_\al^{\eta/N}=E_1(\frac{\eta}{N}+\om_\al)-E_1(\om_\al)\,.
 }
 \end{array}
 \eq
 are related to $R$-matrices multiplied by
permutation operator in "matrix" space ($R^\hbar_{12}(z)\otimes \ti
P_{\ti 1,\ti 2}$). Such object has certainly no classical expansion
beginning with identity matrix. However additional reductions
(\ref{w70}) allow to keep the Lax equations.

The most symmetric form (with respect to $z$ and $\eta$) comes from
another example of the $R$-matrix (\ref{w17}) - the so-called
symmetric $R$-matrix \cite{LOZ162}. It turns into $R_{12}^\hbar(z)$
when $M=1$ and to $R_{12}^z(\hbar)$ when $N=1$. Moreover, such
$R$-matrices satisfy the associative Yang-Baxter equation in the
form (\ref{w18}), but again have wrong local expansion (\ref{w19})
near $z=0$ or $\hbar=0$. Our aim is to describe an integrable Lax
pairs based on the symmetric $R$-matrix, and keep possible symmetry
between a pair of parameters $z$, $\eta$ implying that any of them
can be chosen to be the spectral. For this purpose we are going to
use well-known \cite{RT} relation (\ref{w1005})
$R_{12}^\hbar(z)P_{12}=R_{12}^{z/N}(N\hbar)$. For the Belavin's
$R$-matrix written as (\ref{w1004}) it provides a set of discrete
Fourier transformations (\ref{e913}) and its degenerations given in
Section \ref{s2}.

Consider the Lax matrix related to symmetric $R$-matrix. It is given
by the matrix-valued function on $\Sigma_{z,\tau}\times
\Sigma_{\eta,\tau}$:
 \beq\label{w401}
 \begin{array}{c}
  \displaystyle{
  L(z,\eta)=\sum\limits_{\al\in\,\mZ_{N}^{\times 2}}\sum\limits_{\tal\in\,\mZ_{M}^{\times 2}}
  A^{\al,\tal}\, \Phi_{\al,\tal}(z,\eta)\in {\rm Mat}(K,\mathbb
  C)\,,
  }
 \end{array}
 \eq
where $N$ and $M$ are coprime and $K$ is arbitrary. It has $M^2$
simple poles in variable $z$ at $-N\tom_\tal$, $\tal\in\mZ^{\times
2}_M$ and $N^2$ simple poles in variable $\eta$ at $-\om_\al$,
$\al\in\mZ^{\times 2}_N$.
The answer is as follows. Let $z$ be the spectral parameter.
Introduce $M$-matrix
 \beq\label{w402}
 \begin{array}{c}
  \displaystyle{
  M(z)=-\sum\limits_{\al\neq 0}
  \sum\limits_{\tal} A^{\al,\tal}\, \Phi_{\al,\tal}(z,0)
  -\sum\limits_{\tal}A^{0,\tal} E_1(z+N\tom_\tal)
  }
 \end{array}
 \eq
The l.h.s. of the Lax equations ${\dot L}(z,\eta)=[L(z,\eta),M(z)]$
is the quasi-periodic function in $z$ but its r.h.s. is not
quasi-periodic due to the property of $E_1$ function (\ref{w841}).
This is why we require the following condition:
 \beq\label{w408}
 \begin{array}{c}
  \displaystyle{
  \sum\limits_{\tal}A^{0,\tal}\sim { 1}_K\,.
  }
 \end{array}
 \eq
Using calculations from Appendix B we get the following equations of
motion:
  \beq\label{w409}
 \begin{array}{c}
  \displaystyle{
  {\dot A}^{\al,\tal}=
 \sum\limits_{\tbe\neq\tal,\ga\neq 0} \left([A^{\ga,\tal},A^{\al-\ga,\tbe}]\,\Phi_{\al-\ga,\tbe-\tal}(0,\eta)
   +[A^{\ga,\tga},A^{\al-\ga,\tal}]\,\Phi_{\ga,\tga-\tal}(0,0)\right)
  }
 \end{array}
 \eq
  $$
    \displaystyle{
 +\sum\limits_{\tga\neq\tal} [A^{0,\tga},A^{\al,\tal}]\left( \exp(2\pi\imath\eta \frac{N\tal_2}{M})E_1(N\tom_{\tga-\tal})
 +\Phi_{\al-\ga,\tbe-\tal}(0,\eta)
   \right)
  }
   $$
   $$
    \displaystyle{
 +\sum\limits_{\ga\neq 0} [A^{\ga,\tal},A^{0,\tal}]\left(
 E_1(\eta)+E_1(\om_\ga)-E_1(\eta+\om_\ga+N\tom_\tal)
   \right)
  }
     $$
     $$
    \displaystyle{
 +\sum\limits_{\ga\neq 0} [A^{\ga,\tal},A^{\al-\ga,\tal}]\left(
 E_1(\om_\ga)-E_1(\om_\ga+\eta)
   \right)\,.
  }
 $$
When $M=1$ all the terms are absent except the last line, which
reproduces (\ref{w721}). The whole system of equations (\ref{w409})
can be considered as generalization of (\ref{w721}) describing $M^2$
interacting  $GL(N)$ matrix tops.

Recall that previously we had some special reduction conditions of
types (\ref{w70})-(\ref{w72}). In fact we have them here as well.
Indeed, we may perform the Fourier transformation over sublattice
$\tal\in\mZ^{\times 2}_M\in\mZ^{\times 2}_{MN}$. Then it easy to see
that our requirement (\ref{w408}) is just the one $A^0\sim 1_K$ from
(\ref{w70}) written in the case $N:=NM$ after Fourier transformation
over $\mZ^{\times 2}_M$. The rest of conditions are hardly visible
from the above calculations, but we know that the do exist. Using
notations of (\ref{w302}), (\ref{w303}) we may write them as
  \beq\label{w4099}
 \begin{array}{c}
  \displaystyle{
  \frac{{\mathcal
  A}^{\al+N\ti\ga}}{\vf_{\al+N\ti\ga}(\frac{\eta}{M},\frac{\om_{\al+N\ti\ga}}{M})}=
  \frac{{\mathcal
  A}^{-\al-N\ti\ga}}{\vf_{-\al-N\ti\ga}(\frac{\eta}{M},\frac{-\om_{\al+N\ti\ga}}{M})}
  }
  \,,\quad { {\mathcal A}}^{\al+N\ti\ga}=
  \frac{1}{M}\sum\limits_{\ti\al\in\,\mZ_{M}^{\times 2}}\ti\ka_{\al+N\ti\ga,\ti\al}^2\,
  { A}^{\al,\ti\al}\,.
 \end{array}
 \eq
In a similar way one can use the matrix (\ref{w401}) as the Lax
matrix with spectral parameter $\eta$. The corresponding model
describes $N^2$ interacting $GL(M)$ matrix tops, and acquires other
set of reduction conditions. Viewed in this way we deal with a
spectral surface in $(\lambda,z,\eta)\in\mC^3$:
 \beq\label{w78}
 \begin{array}{c}
\det(\lambda-L(z,\eta))=0
\\
\hbox{reductions of type
1.}\swarrow\quad\quad\quad\quad\quad\searrow\hbox{reductions  of
type 2.}
\\
\det(\lambda-L^\eta(z))=0\,,\quad M=M(z)
\quad\quad\quad\quad\quad\quad\quad \det(\lambda-\ti
L^z(\eta))=0\,,\quad M=\ti M(\eta)
 \end{array}
 \eq
 Integrable cases appear after we choose
a parameter (say $z$) be the spectral one, and fix the parameter
$\eta=const$. Then, with additional reduction constraints such slice
gives a spectral curve of integrable system (\ref{w409}).

In the end let us remark again that matrix (\ref{w401}) bas $M^2$
simple poles in variable $z$ at $-N\tom_\tal$, $\tal\in\mZ^{\times
2}_M$ and $N^2$ simple poles in variable $\eta$ at $-\om_\al$,
$\al\in\mZ^{\times 2}_N$. The described model in both case of
(\ref{w78}) is a matrix extension of the Gaudin-like model. When $z$
is spectral parameter indices $\mZ^{\times 2}_M$ numerates the
marked points while $\mZ^{\times 2}_N$ - are the matrix indices.
Contrariwise, when $\eta$ is the spectral parameter, then
$\mZ^{\times 2}_N$ numerates the marked points while $\mZ^{\times
2}_M$ - are the matrix indices.
A link between Gaudin type models (and/or integrable spin chains)
which interchanges the indices counting marked points and the matrix
indices is known as the spectral duality \cite{Harnad1} (reational
Gaudin -- rational Gaudin), \cite{agt2} (trigonometric Gaudin -- XXX
spin chain), \cite{agt3} (XXZ spin chain -- XXZ spin
chain)\footnote{The type of relation we are discussing here was
first observed for the Toda chain \cite{FT}.}. In our case there is
no direct relation between these two models since the reductions in
(\ref{w78}) are different. Such relation exists at the level of the
off-shell matrix (\ref{w401}) only, where it is given by the
discrete Fourier transformation.

 \section{Appendix A: elliptic functions}
 \def\theequation{A.\arabic{equation}}
 \setcounter{equation}{0}

The Kronecker function on elliptic curve $\mC/(\mZ\oplus\tau\mZ)$
with moduli $\tau$ (Im$\tau>0$):
  \beq\label{w80}
  \begin{array}{l}
  \displaystyle{
 \phi(\eta,z)=\frac{\vth'(0)\vth(\eta+z)}{\vth(\eta)\vth(z)}\,,\quad\quad
\vth(z)=\displaystyle{\sum _{k\in \mathbb Z}} \exp \left ( \pi
\imath \tau (k+\frac{1}{2})^2 +2\pi \imath
(z+\frac{1}{2})(k+\frac{1}{2})\right )\,.
 }
 \end{array}
 \eq
Next function is its derivative:
  \beq\label{w81}
  \begin{array}{l}
  \displaystyle{
 f(z,u)\equiv\p_u\phi(z,u)=\phi(z,u)(E_1(z+u)-E_1(u))\,,
 }
 \end{array}
 \eq
where the first Eisenstein function is used:
  \beq\label{w82}
  \begin{array}{c}
  \displaystyle{
 E_1(z)=\vth'(z)/\vth(z)\,,\quad\quad E_1(-z)=-E_1(z)\,.
 }
 \end{array}
 \eq
The second Eisenstein function:
   \beq\label{w83}
  \begin{array}{c}
  \displaystyle{
   E_2(z)= -\p_z E_1(z)\,,\quad\quad E_2(z)=E_2(-z)
 }
 \end{array}
 \eq
 The functions (\ref{w82}), (\ref{w83}) are simply related to the
 Weierstrass $\zeta$- and $\wp$-functions:
   \beq\label{w84}
  \begin{array}{c}
  \displaystyle{
   E_1(z)=\zeta(z)+\frac{z}{3}\frac{\vth'''(0)}{\vth'(0)}\,,\quad\quad
   E_2(z)= \wp(z)-\frac{1}{3}\frac{\vth'''(0)}{\vth'(0)}\,.
 }
 \end{array}
 \eq
Quasiperiodic properties:
 \beq\label{w841}
 \begin{array}{c}
  \displaystyle{
\phi(x+1,y)=\phi(x,y)\,,\qquad \phi(x+\tau,y)=\exp(-2\pi\imath
y)\phi(x,y) }
 \\ \ \\
   \displaystyle{
   E_1(z+1)=E_1(z)\,,\quad E_1(z+\tau)=E_1(z)-2\pi\imath\,,\quad
   E_2(z+1)=E_2(z+\tau)=E_2(z)\,.
  }
 \end{array}
 \eq
The Fay identity:
  \beq\label{w85}
  \begin{array}{c}
  \displaystyle{
\phi(z,q)\phi(w,u)=\phi(z-w,q)\phi(w,q+u)+\phi(w-z,u)\phi(z,q+u)
 }
 \end{array}
 \eq
 Its degenerations:
  \beq\label{w86}
  \begin{array}{c}
  \displaystyle{
 \phi(z,q)\phi(w,q)=\phi(z+w,q)(E_1(z)+E_1(w)+E_1(q)-E_1(z+w+q))\,,
 }
 \end{array}
 \eq
  \beq\label{w87}
  \begin{array}{c}
  \displaystyle{
 \phi(z,x)f(z,y)-\phi(z,y)f(z,x)=\phi(z,x+y)(\wp(x)-\wp(y))\,.
 }
 \end{array}
 \eq
The following set of $N^2$ functions is widely used:
 \beq\label{w88}
 \begin{array}{c}
  \displaystyle{
 \vf_\al(z,\eta+\om_\al)=\exp(2\pi\imath\,z\,\p_\tau\om_\al)\,\phi(z,\eta+\om_\al)\,,
 }
 \end{array}
 \eq
where
 \beq\label{w89}
 \begin{array}{c}
  \displaystyle{
\om_\al=\frac{\al_1+\al_2\tau}{N}\,,\quad
 \p_\tau\om_\al=\frac{\al_2}{N}\,,\quad
 \al=(\al_1,\al_2)\in \mZ_N\times\mZ_N\,.
 }
 \end{array}
 \eq
 The index $\al$ in $\vf_\al$ stands for the exponential factor
depending on $\al$ and the first argument. Similarly,
 \beq\label{w90}
 \begin{array}{c}
  \displaystyle{
 f_\al(z,\om_\al)=\exp(2\pi\imath\,z\,\p_\tau\om_\al)\,f(z,\om_\al)\,,
 \quad\quad
 (\al_1,\al_2)\neq (0,0)\,.
 }
 \end{array}
 \eq
Plugging (\ref{w88}) into (\ref{w85}), (\ref{w86}) and (\ref{w87})
we get:
 \beq\label{w91}
 \begin{array}{l}
  \displaystyle{
 \vf_\be(x,\eta+\om_\be)\vf_\ga(y,\om_\ga)=
 }
 \\ \ \\
  \displaystyle{
 =\vf_\be(x-y,\eta+\om_\be)\vf_{\be+\ga}(y,\eta+\om_{\be+\ga})+
 \vf_\ga(y-x,\om_\ga)\vf_{\be+\ga}(x,\eta+\om_{\be+\ga})\,,
 }
 \end{array}
 \eq
 \beq\label{w92}
 \begin{array}{l}
  \displaystyle{
  \vf_\be(z,\eta+\om_\be)\vf_\ga(z,\om_\ga)=
 }
 \\ \ \\
  \displaystyle{
 =\vf_{\be+\ga}(z,\eta+\om_{\be+\ga})\,(E_1(z)+E_1(\eta+\om_\be) +E_1(\om_\ga)-E_1(z+\eta+\om_{\be+\ga}))
 }
 \end{array}
 \eq
and
 \beq\label{w93}
 \begin{array}{c}
  \displaystyle{
 \vf_\be(z,\om_\be)f_\ga(z,\om_\ga)-\vf_\ga(z,\om_\ga)f_\be(z,\om_\be)=
 \vf_{\be+\ga}(z,\om_{\be+\ga})(\wp(\om_\be)-\wp(\om_{\ga}))\,.
 }
 \end{array}
 \eq

 \section{Appendix B}
 \def\theequation{B.\arabic{equation}}
 \setcounter{equation}{0}

Consider the r.h.s. of the Lax equation ${\dot
L}(z,\eta)=[L(z,\eta),M(z)]$ for the Lax pair (\ref{w401}),
(\ref{w402}):
 \beq\label{w403}
 \begin{array}{c}
  \displaystyle{
  [L(z,\eta),M(z)]=
  }
 \end{array}
 \eq
$$
\sum\limits_{\be,\tbe,\tga}\sum\limits_{\ga\neq
  0}\, [A^{\ga,\tga},A^{\be,\tbe}]\, \Phi_{\be,\tbe}(z,\eta)\,\Phi_{\ga,\tga}(z,0)
  +\sum\limits_{\be,\tbe,\tga}\,
  [A^{0,\tga},A^{\be,\tbe}]\, \Phi_{\be,\tbe}(z,\eta)\,E_1(z+N\tom_\tga)
$$
Both sums are subdivided into two parts: with $\tbe=\tga$ and
$\tbe\neq\tga$:
 \beq\label{w404}
 \begin{array}{c}
  \displaystyle{
  [L(z,\eta),M(z)]=
  }
 \end{array}
 \eq
$$
\sum\limits_{\be,\tbe\neq\tga}\left(\sum\limits_{\ga\neq
  0}\, [A^{\ga,\tga},A^{\be,\tbe}]\, \Phi_{\be,\tbe}(z,\eta)\,\Phi_{\ga,\tga}(z,0)
  +
  [A^{0,\tga},A^{\be,\tbe}]\,
  \Phi_{\be,\tbe}(z,\eta)\,E_1(z+N\tom_\tga)\right)
$$
$$
+\sum\limits_{\be,\tbe}\left(\sum\limits_{\ga\neq
  0}\, [A^{\ga,\tbe},A^{\be,\tbe}]\, \Phi_{\be,\tbe}(z,\eta)\,\Phi_{\ga,\tbe}(z,0)
  +
  [A^{0,\tbe},A^{\be,\tbe}]\,
  \Phi_{\be,\tbe}(z,\eta)\,E_1(z+N\tom_\tbe)\right)
$$
In the second sum in the last line we rename $\be$ into $\ga$
(mention that the terms with $\be=0$ are cancelled out). The first
sum in the last line is subdivided into two parts: with $\be=0$ and
$\be\neq 0$:
 \beq\label{w405}
 \begin{array}{c}
  \displaystyle{
  [L(z,\eta),M(z)]=
  }
 \end{array}
 \eq
$$
\sum\limits_{\be,\tbe\neq\tga}\left(\sum\limits_{\ga\neq
  0}\, [A^{\ga,\tga},A^{\be,\tbe}]\, \Phi_{\be,\tbe}(z,\eta)\,\Phi_{\ga,\tga}(z,0)
  +
  [A^{0,\tga},A^{\be,\tbe}]\,
  \Phi_{\be,\tbe}(z,\eta)\,E_1(z+N\tom_\tga)\right)
$$
$$
+\sum\limits_{\tbe}\sum\limits_{\ga\neq
  0}\left( [A^{\ga,\tbe},A^{0,\tbe}]\, \Phi_{0,\tbe}(z,\eta)\,\Phi_{\ga,\tbe}(z,0)
  +
  [A^{0,\tbe},A^{\ga,\tbe}]\,
  \Phi_{\ga,\tbe}(z,\eta)\,E_1(z+N\tom_\tbe)\right)
$$
$$
+\sum\limits_{\tbe}\sum\limits_{\be\neq0,\ga\neq
  0}\, [A^{\ga,\tbe},A^{\be,\tbe}]\,
  \Phi_{\be,\tbe}(z,\eta)\,\Phi_{\ga,\tbe}(z,0)\,.
$$
Consider the r.h.s. of  (\ref{w405})  in more detail. The first sum
in its first line is simplified by (\ref{w33}), and we leave the
second term in the first line as is. The second line is simplified
via (\ref{w34}). The last line is also modified through (\ref{w34})
likewise it was made in (\ref{w13}):
 \beq\label{w406}
 \begin{array}{c}
  \displaystyle{
  \sum\limits_{\tbe}\sum\limits_{\be\neq0,\ga\neq
  0}\, [A^{\ga,\tbe},A^{\be,\tbe}]\,
  \Phi_{\be,\tbe}(z,\eta)\,\Phi_{\ga,\tbe}(z,0)=
  }
  \\ \ \\
    \displaystyle{
  =\frac12\sum\limits_{\tbe}\sum\limits_{\be\neq0,\ga\neq
  0}\, [A^{\ga,\tbe},A^{\be,\tbe}]\,
  \left(\Phi_{\be,\tbe}(z,\eta)\,\Phi_{\ga,\tbe}(z,0)-\Phi_{\ga,\tbe}(z,\eta)\,\Phi_{\be,\tbe}(z,0)\right)\stackrel{(\ref{w34})}{=}
  }
 \end{array}
 \eq
     $$
  \frac12\sum\limits_{\tbe}\sum\limits_{\be\neq0,\ga\neq
  0}\, [A^{\ga,\tbe},A^{\be,\tbe}]\,
  \Phi_{\be+\ga,\tbe}(z,\eta)
  (E_1(\eta+\om_\be)+E_1(\om_\ga)-E_1(\eta+\om_\ga)-E_1(\om_\be))=
  $$
      $$
  =\sum\limits_{\tbe}\sum\limits_{\be\neq0,\ga\neq
  0}\, [A^{\ga,\tbe},A^{\be,\tbe}]\,
  \Phi_{\be+\ga,\tbe}(z,\eta)
  (E_1(\eta+\om_\be)-E_1(\om_\be))\,.
  $$
Finally, we have
 \beq\label{w407}
 \begin{array}{c}
  \displaystyle{
  [L(z,\eta),M(z)]=
  }
 \end{array}
 \eq
$$
=\sum\limits_{\be,\tbe\neq\tga}\sum\limits_{\ga\neq
  0}\, [A^{\ga,\tga},A^{\be,\tbe}]\,
   \left(\Phi_{\be,\tbe-\tga}(0,\eta)\,\Phi_{\be+\ga,\tga}(z,\eta)+\Phi_{\ga,\tga-\tbe}(0,0)\,\Phi_{\be+\ga,\tbe}(z,\eta)
  \right)
$$
$$
+\sum\limits_{\be,\tbe\neq\tga}
  [A^{0,\tga},A^{\be,\tbe}]\,
  \Phi_{\be,\tbe}(z,\eta)\,E_1(z+N\tom_\tga)
$$
$$
+\sum\limits_{\tbe}\sum\limits_{\ga\neq
  0} [A^{\ga,\tbe},A^{0,\tbe}]\,
  \Phi_{\ga,\tbe}(z,\eta) \left( E_1(\eta)+E_1(\om_\ga)-E_1(z+\eta+\om_{\ga}+N\tom_\tbe)\right)
$$
$$
+\sum\limits_{\tbe}\sum\limits_{\be\neq0,\ga\neq
  0}\, [A^{\ga,\tbe},A^{\be,\tbe}]\,
  \Phi_{\be+\ga,\tbe}(z,\eta)
  (E_1(\eta+\om_\be)-E_1(\om_\be))\,.
$$
The r.h.s. has simple poles at $z=-N\tom_\tal$,
$\tal\in\,\mZ_{M}^{\times 2}$ (apparent pole at
$z=-\eta-\om_{\ga}-N\tom_\tbe$ in the third line is cancelled by the
zero of $\Phi_{\ga,\tbe}(z,\eta)$). In order to get equations of
motion we are going to compare its residues with those of functions
$\Phi_{\ga,\tbe}(z,\eta)$ entering ${\dot L}(z,\eta)$. Recall that
$\res_{z=-N\tom_\tal} \Phi_{\al,\tal}(z,\eta)=\exp(2\pi\imath \eta
N{\tal_2}/{M})$. This gives the equations of motion (\ref{w409}).

\subsubsection*{Acknowledgments}
The work was performed at the Steklov Mathematical Institute of
Russian Academy of Sciences, Moscow. It was supported by Russian
Science Foundation (RSCF) grant 14-50-00005.

\begin{small}

\end{small}

\end{document}